# Three-step growth of vapor-deposited ice under mesospheric temperature and water vapor conditions

*Reo Sato, Kentaro Noguchi, Hiroyuki Koshida, Atsuki Ishibashi, Tetsuya Hama*[*]

Komaba Institute for Science and Department of Basic Science,

The University of Tokyo, Meguro, Tokyo 153-8902, Japan

Correspondence to: Tetsuya Hama

E-mail: hamatetsuya@g.ecc.u-tokyo.ac.jp    Phone: +81-3-5452-6288

TOC GRAPHICS

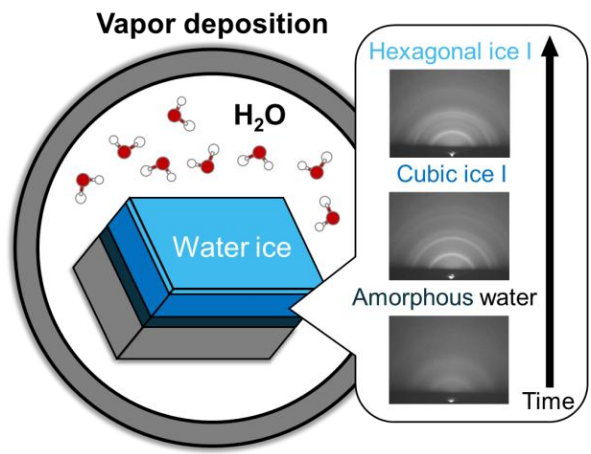

ABSTRACT.

Polar mesospheric clouds provide clues to physicochemical processes in the mesosphere and lower thermosphere. However, the heterogeneous nucleation and growth processes of water ice under polar mesospheric conditions are poorly understood, especially at the nanoscale. This study used reflection high-energy electron diffraction and infrared reflection–absorption spectroscopy to analyze the structure of vapor-deposited ice at polar mesospheric temperature (120 K) under vapor pressure ($10^{-6}$ Pa) conditions. The ice appeared to grow in three steps during vapor deposition, being amorphous water for the first 15 nm, then cubic ice up to 50 nm, and finally hexagonal ice subsequently. This three-step growth suggests that the three observed phases can coexist in polar mesospheric clouds, depending on the thickness of water ice. The finding of the three-step growth also shows that the Ostwald rule of stages can hold for vapor-deposited ice at low temperature.

Polar mesospheric clouds, also known as noctilucent clouds, are the highest clouds in Earth's atmosphere. They occur within the summer mesopause region at mid and high latitudes at altitudes of about 83 km.[1–3] Satellite observations of these clouds indicate they are composed mainly of nanometer-sized water ice particles with small amounts of meteoric smoke (0.01%–3.00% by volume).[4–7] The formation mechanism of polar mesospheric clouds is a topic of intense research because they can be used to study the physics and chemistry of the mesosphere and lower thermosphere region; they are also potential indicators of global climate change.[1–3,8]

Both heterogeneous nucleation (i.e., the deposition of water vapor on nuclei such as dust grains) and homogeneous nucleation (i.e., the formation of new water nuclei directly from the gas phase) have been proposed to explain the formation of water ice in polar mesospheric clouds.[1,9,10] A recent theoretical study reported that homogeneous nucleation is unlikely to occur in the mesosphere, whereas heterogeneous nucleation occurs effectively.[11] Nanometer-sized meteoric smoke particles can serve as nuclei for heterogeneous nucleation in the mesosphere.[1,2,8,11,12] As a heterogeneous nucleation system, water vapor deposition on a cold substrate under vacuum has long been studied in the laboratory.[13–17] At low temperatures relevant to the formation of polar mesospheric clouds (below about 145 K),[8] three forms of ice I (stable hexagonal ice $I_h$, metastable cubic ice $I_c$, and stacking disordered ice $I_{sd}$ with disordered hexagonal and cubic stacking sequences)[18] and amorphous water without a long-range ordered structure can be formed by vapor deposition on a substrate.[13,14,19–22] As the structure of water ice can influence its macroscopic properties such as density,[20,23] thermal conductivity,[24–28] vapor pressure over the solid,[29–32] and particle shape,[33,34] studying the structure of ice formed under polar mesospheric conditions is critical to understanding the role of polar mesospheric clouds in atmospheric processes.

It is often implicitly assumed that the structure of vapor-deposited ice is determined solely by the substrate temperature.[19,21,22] For example, Mangan et al. used X-ray diffraction to study the structure of vapor-deposited ice formed between 88 and 145 K at a total pressure of 1000–2000 Pa (with a partial vapor pressure of approximately 0.3 Pa), finding that amorphous water, ice $I_{sd}$, and ice $I_h$ formed at substrate temperatures of 88–120, 121–135, and 140–145 K, respectively.[19] Using reflection high-energy electron diffraction (RHEED), Kouchi et al. demonstrated that the crystallinity of vapor-deposited ices formed at 95–110 K is determined not only by the substrate temperature, but also by the water vapor pressure in the range of $10^{-5}$ to $10^{-3}$ Pa.[35] Considering the typical partial water vapor pressure in the mesosphere ($10^{-7}$–$10^{-5}$ Pa),[36] it is necessary to conduct vapor deposition experiments at a water vapor pressure below $10^{-5}$ Pa.

In addition, the effect of the size (thickness) of vapor-deposited ices on their structure should be investigated at the nanoscale to aid our understanding of the ice structures in polar mesospheric clouds, which are typically less than about 100 nm in size.[3,6,37] Homogeneous ice nucleation experiments and theoretical calculations have identified the dependence of the stable configuration of water ice on particle size.[38–40] Moberg et al. measured the infrared (IR) spectra of size-selected water clusters $(H_2O)_n$ around 150 K, finding that the lowest-energy structures are amorphous for $n$ values below about 70, while clusters with $n = 90$ can have ice I structures.[39] In contrast to recent progress in the study of homogeneous ice nucleation, our understanding of heterogeneously nucleated ice remains limited.

The present study developed a new apparatus for the *in situ* structural analysis of vapor-deposited ice using a combination of infrared reflection-absorption spectroscopy (IRRAS) and RHEED (Fig. 1; for details, see the Experimental Methods), and investigated the structure of vapor-deposited ice formed under polar mesospheric conditions (i.e., a substrate temperature of 120 K and water vapor pressure of $10^{-6}$ Pa). IRRAS can provide information on the bulk structure of vapor-deposited ice and quantify its thickness.[41–45] RHEED can probe the top several layers of ice samples and distinguish among amorphous water, ice $I_c$, and ice $I_h$.[35,46–49] Despite the constant temperature and vapor pressure condition, combined IRRAS and RHEED analysis revealed that amorphous water, ice $I_c$, and ice $I_h$ are sequentially produced on the surface during vapor deposition; i.e., the structure of newly formed ice upon the deposition of water molecules depends on the thickness of the preexisting ice at the nanoscale.

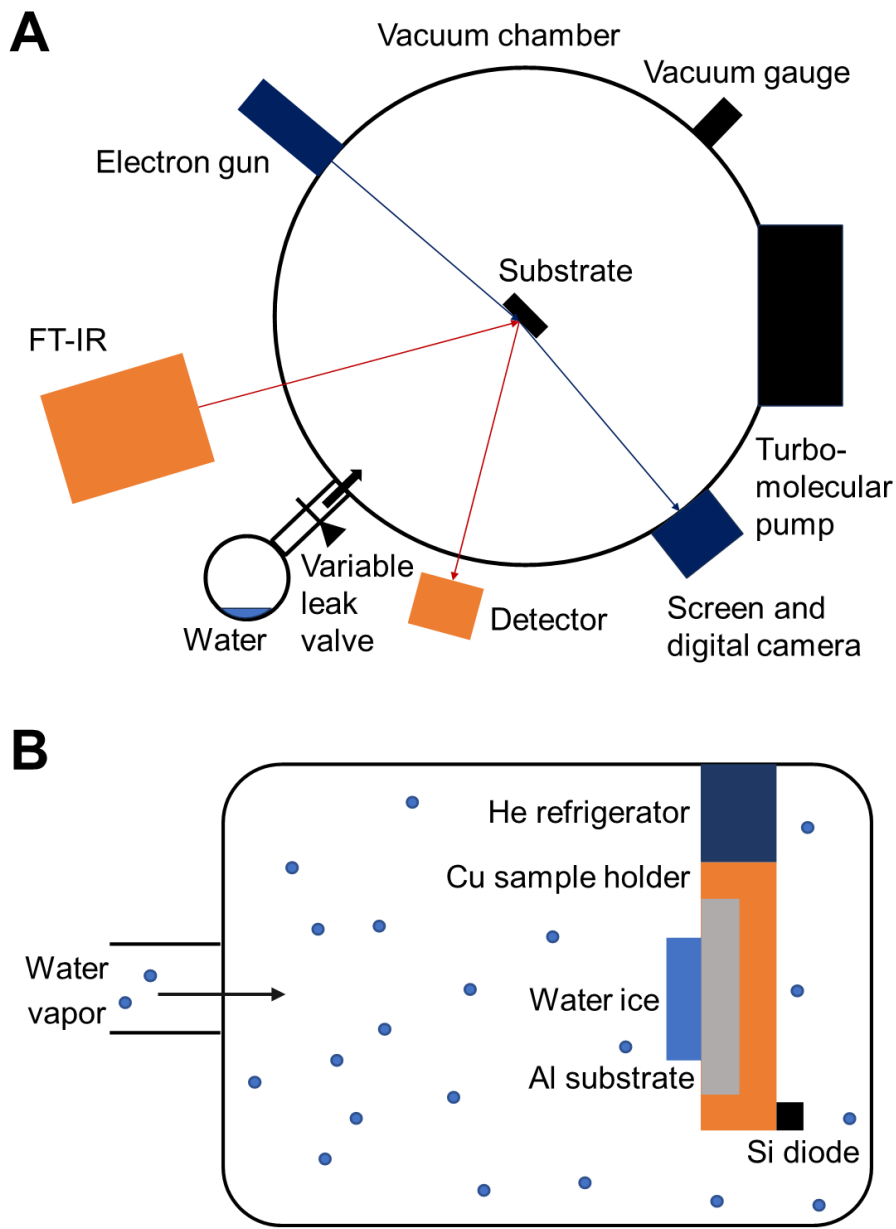

**Figure 1.** Schematic illustration of equipment for RHHED and IRRAS measurements of vapor-deposited ice: (A) overview and (B) close-up of an Al substrate. FT-IR, Fourier transform infrared spectrometer.

Figure 2A show the IRRAS spectra of vapor-deposited ice on an Al substrate at 120 K as a function of deposition time for durations between 30 and 120 min. Analytical expressions for the reflection–absorbance of a thin sample measured by IRRAS using a metallic substrate are summarized elsewhere.[41–45] In brief, Eqs. 1 and 2 describe the reflection–absorbance when using p-polarized IR light ($A_p$) with an angle of incidence $\varphi$ of less than 85°:

$$A_p = -\log_{10}\frac{S^s}{S^b} = \frac{8\pi l \nu}{\ln 10}\frac{n_v^3 \sin^2\varphi}{\cos\varphi} f_{LO}, \tag{1}$$

$$f_{LO} = \mathrm{Im}\left[-\frac{1}{\bar{\varepsilon}_z}\right] = \mathrm{Im}\left[\frac{1}{(n_z+ik_z)^2}\right] = \frac{2n_z k_z}{(n_z^2+k_z^2)^2} \tag{2}$$

where $S$ is the intensity of the reflected IR light of wavenumber $\nu$ (cm$^{-1}$) (i.e., single-beam spectrum); the superscripts $s$ and $b$ indicate sample and background measurements, respectively; and $l$ is the sample thickness. The refractive index of the vacuum is $n_v = 1$. The longitudinal optic energy-loss function of a thin sample, $f_{LO}$, is expressed using the $z$-component (surface-perpendicular component) of the complex permittivity ($\bar{\varepsilon}_z$) or complex refractive index ($n_z + ik_z$) of the thin sample, where $\bar{\varepsilon} = (n + ik)^2$. Equations 1 and 2 mean that the band shape of an IRRAS spectrum is given by $\nu f_{LO}$ and that the sample thickness can be obtained by fitting the IRRAS spectrum with $l$ as the parameter. In this study, we calculated $f_{LO}$ using $n$ and $k$ values reported for amorphous water at 70 K (see also Table S1 and Fig. S1 in the Supporting Information).[50] We assumed $n_z = n$ and $k_z = k$ considering the isotropic orientation of OH groups in amorphous water.[45,51] As negligible reflection–absorbance is induced by s-polarized IR light,[42] a reflection–absorption measurement using unpolarized light gives about half the absorbance ($A_p/2$) as that using only p-polarized light, assuming identical intensities of p- and s-polarized light.

The dashed red lines in Fig. 2A show the calculated spectra ($A_p/2$) assuming thicknesses of $l$ = 5.6, 10, 13, and 17 nm for the vapor-deposited ices prepared after gas exposure for 30, 60, 90, and 120 min, respectively. The band shapes of the IRRAS spectra are well reproduced by Eq. 1 when adopting the $n$ and $k$ values for amorphous water. The growth rate was estimated to be about 0.15 ± 0.01 nm min$^{-1}$ by linear fitting of the ice thicknesses estimated from the IRRAS spectra (Fig. 2B). This corresponds to 0.41 ± 0.03 bilayer min$^{-1}$, assuming that the interlayer spacing between two hexagonal (0001) planes of ice (0.366 nm) corresponds to the thickness of one bilayer of crystalline ice.[52–54] The interlayer spacing is also a reasonable approximation for the single-bilayer thickness of amorphous water.[52] Considering a surface coverage of about 1.1 × 10$^{15}$ molecule cm$^{-2}$ on ice I$_h$(0001),[55,56] the gas flux ($f$) during vapor deposition was also estimated to be 7.8 ± 0.5 × 10$^{12}$ cm$^{-2}$ from the growth rate. This corresponds to the deposition pressure of $p$ = 2.2 ± 0.2 × 10$^{-6}$ Pa from the relationship $p = f\sqrt{2\pi m k_B T}$, where $m$, $k_B$, and $T$ represent the mass of water molecules (2.99 × 10$^{-26}$ kg), the Boltzmann constant (1.38 × 10$^{-23}$ J K$^{-1}$), and the gas temperature (295 K), respectively.[57,58] Hence, our experimental conditions simulated the partial water vapor pressure in the polar mesosphere (10$^{-7}$–10$^{-5}$ Pa).[36] At 120 K, the probability of gaseous water molecules sticking to an ice surface is almost unity, regardless of the ice structure (i.e., amorphous or crystalline),[20,59–61] and the desorption of adsorbed water molecules from the ice surface is negligible.[29,31,57,62]

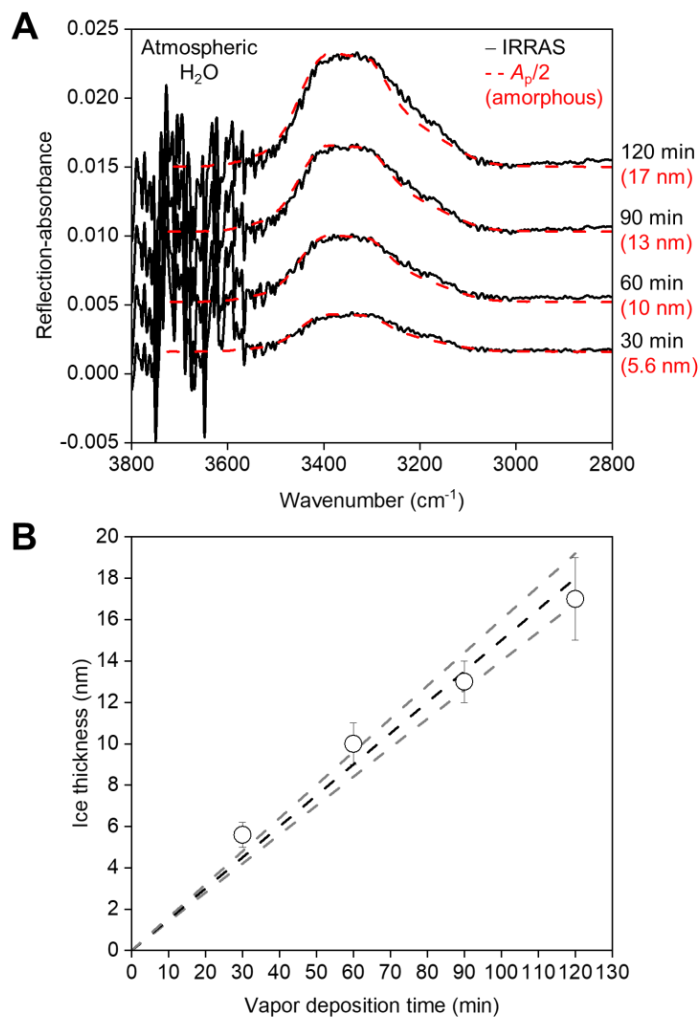

**Figure 2.** (A) IRRAS spectra in the region of the OH stretching vibration (3800–2800 cm$^{-1}$) of vapor-deposited ice at 120 K for 30–120 min at $2.2 \times 10^{-6}$ Pa. Dashed red lines are spectra ($A_p^{RA}/2$ in eq 1) calculated for amorphous water assuming ice thicknesses of $l$ = 5.6, 10, 13, and 17 nm. The noise peaks at 3800–3600 cm$^{-1}$ are attributable to atmospheric water vapor in the IR beam path outside the chamber. Data are offset for clarity. (B) Ice thickness with respect to vapor deposition time (open circles). Dashed lines are linear fitting assuming deposition at 0.015 (black) and 0.014 and 0.016 nm min$^{-1}$ (gray).

Figure 3A–D shows the RHEED patterns of vapor-deposited ices after deposition for 30, 60, 90, and 120 min, respectively. Vague halo patterns in Fig. 3A indicate the formation of amorphous water after deposition for 30 min, consistent with the IRRAS spectrum (Fig. 2A). Increasing the deposition time to 60–120 min ($l$ = 10–17 nm) increased the image contrast, and distinct Debye–Scherrer rings appeared, which were indexed as the (111), (220), (311), (331), and (422) diffractions of ice $I_c$ (Fig. 3B–D).[24,35,46–48,63,64] As the corresponding IRRAS spectra (Fig. 2A) still exhibit features of amorphous water, the bulk ice structure appeared to be predominantly amorphous. Hence, the appearance of the Debye–Scherrer rings indicated by RHEED suggests that many small crystals of ice $I_c$ started to form on the surface of amorphous water at $l$ = 10–17 nm.[49,65] As no hexagonal diffraction is apparent in Fig. 3,[46,47] the extent of stacking disorder in the vapor-deposited ices was below the detection limit of the RHEED experiments. For reference, the preparation of a large amount of pure ice $I_c$ without stacking disorder for neutron and X-ray diffraction experiments requires a starting material prepared at high pressure; e.g., Del Rosso et al.[66,67] employed hydrogen-filled ice in the $C_0$ phase prepared at a pressure above 430 MPa (at 255 K), and Komatsu et al.[68] used $C_2$ hydrogen hydrate prepared at 3 GPa and room temperature.

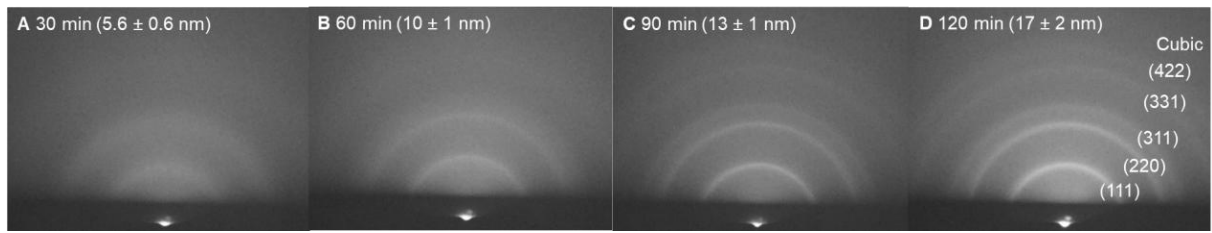

**Figure 3.** RHEED patterns of vapor-deposited ice at 120 K as a function of deposition time at and $2.2 \times 10^{-6}$ Pa for (A) 30, (B), 60, (C) 90, and (D) 120 min. Labels (*hkl*) are for the planes of cubic ice.

The structure of the ice further changed during its prolonged growth by vapor deposition for 420 min. The black solid lines in Fig. 4A show the IRRAS spectra of ices deposited at 120 K for various durations between 120 and 420 min. The thickness of ice after deposition for 420 min was estimated to be 63 ± 4 nm from the growth rate of 0.15 ± 0.01 nm min$^{-1}$. The figure also shows calculated spectra (dashed red lines) for amorphous water ($A_p/2$ in Eq. 1) assuming thicknesses ranging from 17 to 63 nm. The shapes of the IRRAS spectra gradually diverge from the calculated spectra as the vapor deposition time increases to 420 min, and a new peak appears at around 3300 cm$^{-1}$. This new peak is close to that for polycrystalline ice I (around 3280 cm$^{-1}$) calculated using Eqs. 1 and 2 with literature values for $n$ and $k$ (see also Table S2 and Fig. S1 in the Supporting Information).[50] Figure 4A shows that crystalline ice grew with $l \geq 17$ nm during vapor deposition for 120–420 min.

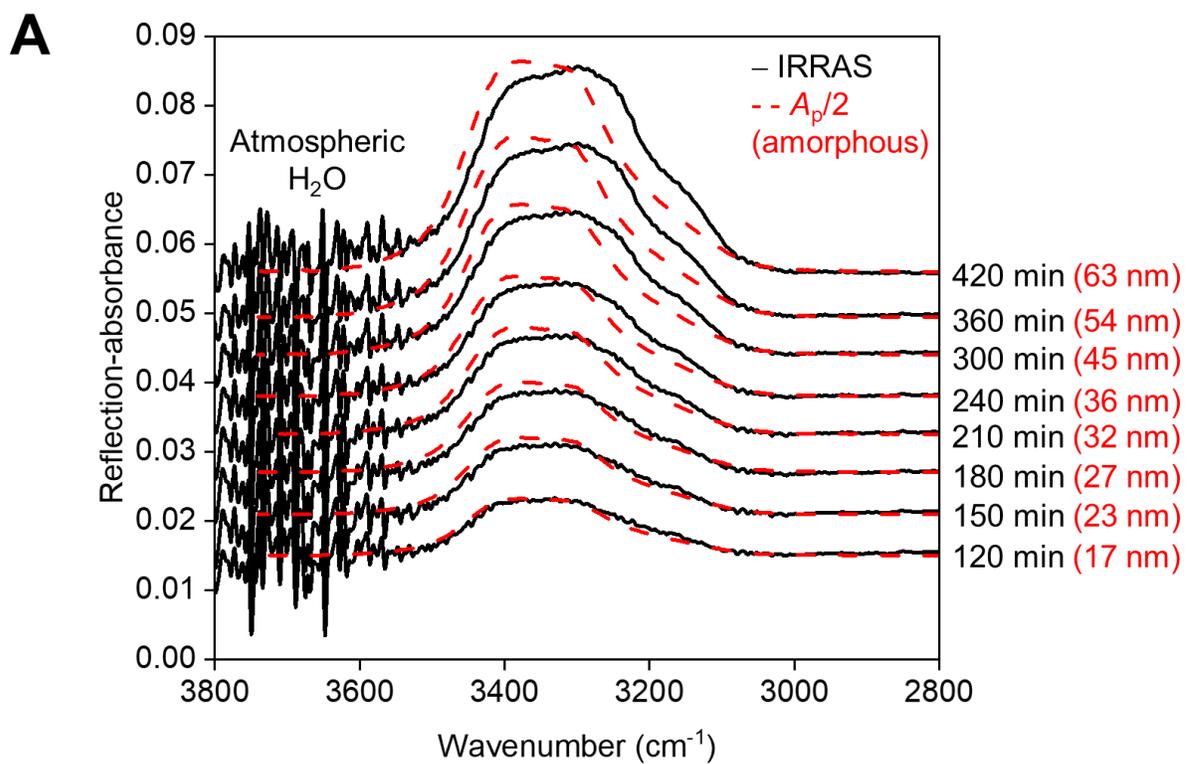
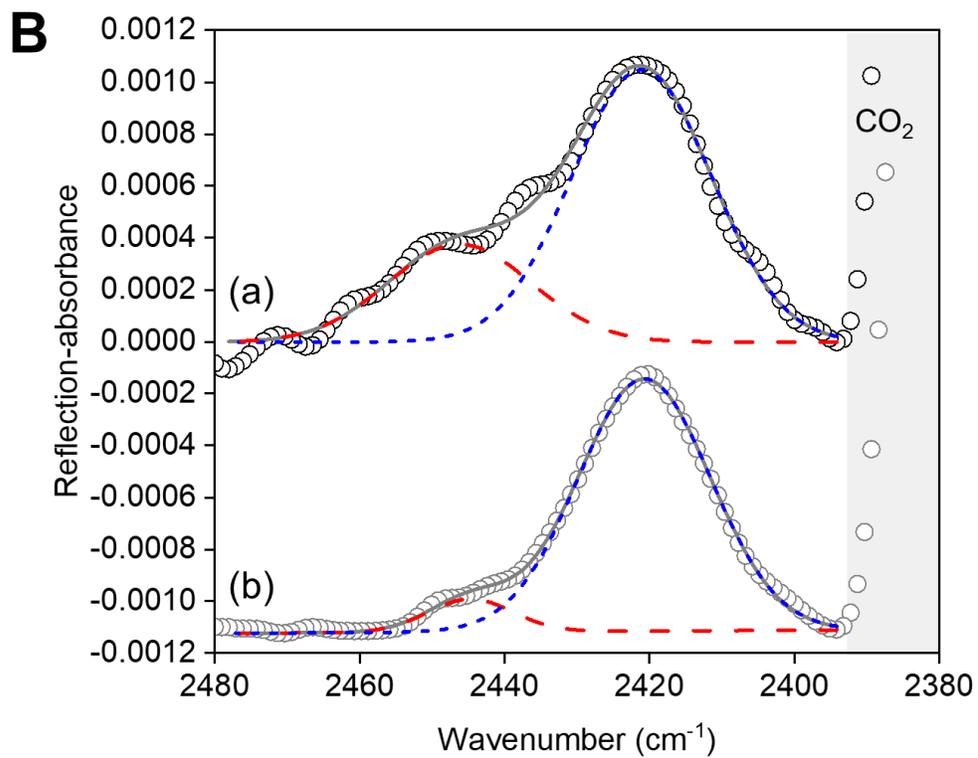

**Figure 4.** IRRAS spectra for vapor-deposited ice at 120 K. (A) OH stretching vibration region at 3800–2800 cm$^{-1}$ after deposition for 120–420 min at 2.2 × 10$^{-6}$ Pa. Dashed red lines show the calculated spectra ($A_p/2$ in Eq. 1) for amorphous water assuming thicknesses of $l$ = 17–63 nm. Noise peaks at 3800–3600 cm$^{-1}$ are attributed to atmospheric water vapor in the IR beam path outside the chamber. (B) Decoupled OD stretching vibration region at 2480–2380 cm$^{-1}$ for (a) vapor-deposited ice after deposition for 420 min at 2.2 × 10$^{-6}$ Pa (black open circles) and (b) reference ice I at 120 K (gray open circles). Reference ice I was prepared at 130 K by vapor deposition for 42 min at about 2.2 × 10$^{-5}$ Pa. For comparison, the vertical scale in (b) was adjusted to match that in (a). In (a) and (b), the experimental data (open circles) were reproduced by Gaussian fitting (gray lines) with two components: one for crystalline ice (blue dotted line) and another amorphous ice (red dashed line) (see also Table S3 in the Supporting Information). Gray regions below about 2390 cm$^{-1}$ were influenced by atmospheric carbon dioxide ($CO_2$) in the IR beam path outside the chamber. The data are offset for clarity.

Given that IR spectroscopy can barely distinguish between ice $I_c$ and ice $I_h$,[69,70] RHEED measurements were also conducted (Fig. 5). The obtained RHEED patterns show that a cubic stacking sequence (ice $I_c$) was dominant during deposition for up to 240 min (Fig. 5A–E), whereas a hexagonal (103) diffraction pattern appeared after 300 min (Fig. 5F). Furthermore, an additional hexagonal (102) diffraction pattern became visible after deposition for 360–420 min (Fig. 5G and H).[46,47] These results indicate that the growing ice formed a hexagonal stacking sequence (ice $I_h$) when it became thicker than 45 ± 3 nm. The present RHEED results do not easily enable an estimation of the extent of stacking disorder (ice $I_{sd}$) on the ice surface, because the diffraction patterns of ice $I_c$ overlap those of ice $I_h$.[47,48]

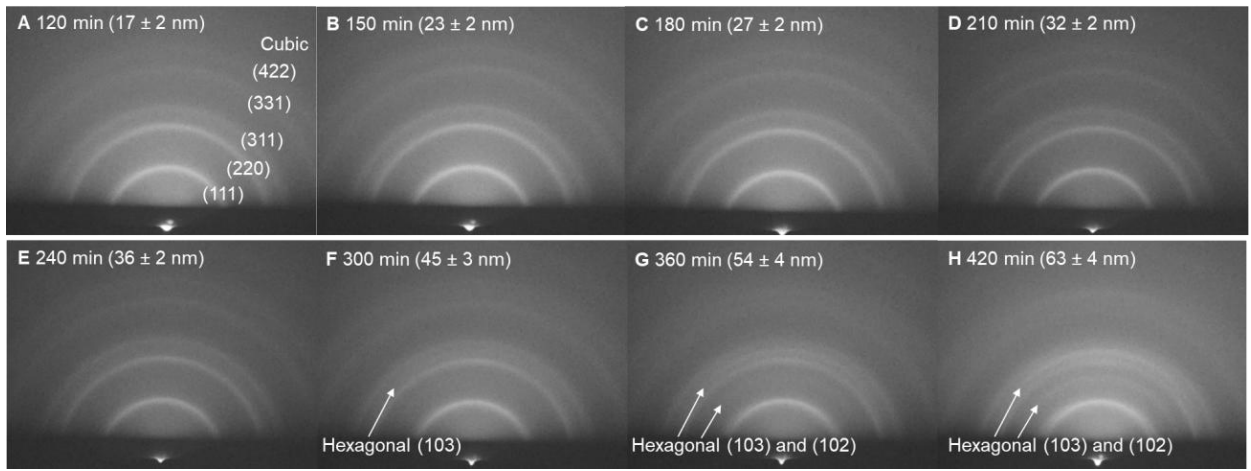

**Figure 5.** RHEED patterns of vapor-deposited ice at 120 K as a function of deposition time and $2.2 \times 10^{-6}$ Pa for (A) 120, (B) 150, (C) 180, (D) 210, (E) 240, (F) 300, (G) 360, and (H) 420 min. Labels (hkl) are for the planes of cubic and hexagonal ices.

Considering the penetration depth of electrons for RHEED (approximately 2–3 nm; for details, see the Experimental Methods), it is not possible to investigate the whole crystallinity of the vapor-deposited ice with $l = 63 \pm 4$ nm (Fig. 5H). In IRRAS, the OH-stretching vibrations in $H_2O$ ice are delocalized by intermolecular vibrational couplings, which complicate interpretation of the vibrational spectra.[71–73] Therefore, the decoupled OD stretching vibration band of dilute semiheavy water (HDO) was used to study the bulk ice structure [Fig. 4B(a)], because its peak wavenumber reflects the local lattice structure (oxygen–oxygen distance) (see also the Experimental Methods).[74–76] We also measured a reference IRRAS spectrum for ice I prepared by vapor deposition [Fig. 4B(b)]: water vapor was first introduced onto the Al substrate at 130 K for 42 min at about $2.2 \pm 0.2 \times 10^{-5}$ Pa to form ice I (Fig. S2). The prepared ice I was then cooled to 120 K for IRRAS measurements. Both IRRAS spectra show a peak at 2421 $cm^{-1}$ that is characteristic of ice $I_h$,[74] whereas the vapor-deposited ice prepared at 120 K shows a stronger shoulder feature at 2480−2430 $cm^{-1}$ compared with the reference ice I. This indicates that amorphous water coexisted with crystalline ice in the bulk.[76] The amorphous proportion was estimated to be $0.30 \pm 0.04$ from Gaussian fitting of the two amorphous and crystalline features in Fig. 4B(a) (see also Table S3 in the Supporting Information). This value is close to that estimated from the ratio of the ice thickness at which ice $I_c$ formation was clearly observed ($l$ = 13–17 nm) to the whole ice thickness ($l$ = 63 nm): i.e., 0.21–0.27 (Figs. 2 and 5). These results suggest that once amorphous water formed, it could not easily attain a crystalline configuration by structural rearrangement in the bulk during further vapor deposition, which in turn means that cubic and hexagonal stacking sequences were produced on the ice surface upon the deposition of water molecules from the gas phase. This inference is reasonable, considering that the timescale required for 100% crystallization of vapor-deposited amorphous water at 120 K is more than $5 \times 10^5$ s (8300 min), which is much longer than the deposition time of 420 min used in this study.[26,27,77,78] The transformation of ice $I_c$ to ice $I_h$ in bulk samples is also negligibly slow at 120 K.[67,77,79–83] We confirmed that amorphous water deposited for 30 min (Fig. 3A) did not crystallize after annealing for 90 min at 120 K (120 min in total; Fig. S3) and that ice $I_c$ deposited for 210 min (Fig. 5D) did not transform into ice $I_h$ after annealing for 210 min at 120 K (420 min in total; Fig. S4).

Vapor-deposited ice has a long history of research, for at least 90 years.[13,14] However, the structural changes in such ice have often been studied during annealing at constant temperature or under heating,[16,78,80,84–86] and the size (thickness) dependence of the ice structure during vapor deposition has received little attention, especially in terms of multilayer coverage at the nanoscale. We first discuss the conditions required for the formation of crystalline and amorphous materials following vapor deposition.[35,76,87–89] Crystalline ice forms when water molecules settle at energetically favorable sites. During vapor deposition, water molecules must be able to diffuse to a suitable site before new monomers are adsorbed and molecular motion ceases. For this to occur, water molecules should diffuse over an area larger than the lattice site area of the crystalline ice ($A_{ice}$, cm$^2$) within the coverage time, $t_{cover}$ (s), which is the time required for adsorbed water molecules to cover the surface during vapor deposition:

$$D_s t_{cover} \geq A_{ice}, \tag{3}$$

where $D_s$ (cm$^2$ s$^{-1}$) is the surface diffusion coefficient of water molecules. In contrast, amorphous water will form if the impinging water molecules readily create hydrogen bonds with the ice surface before they can diffuse to the energetically favorable crystalline sites:[35,47,76]

$$D_s t_{cover} < A_{ice}, \tag{4}$$

Area $A_{ice}$ is estimated as the inverse of the surface number density on ice $I_h(0001)$ and $I_c(111)$; i.e., $A_{ice} = 1/1.1 \times 10^{15} = 8.8 \times 10^{-16}$ cm$^2$.[55,56] Time $t_{cover}$ is taken to be 147 ± 10 s from the ice growth rate (0.41 ± 0.03 bilayer min$^{-1}$). Equations 3 and 4 are valid when the effect of heating due to the latent heat of condensation is negligible.[24,35,90–95] We confirmed that increasing the deposition pressure by one order of magnitude (i.e., to about 2.2 × 10$^{-5}$ Pa) caused only amorphous water to form at 120 K, even after deposition for 120 min (Fig. S5). Hence, neither the latent heat caused by vapor deposition nor the ambient room-temperature radiation from the inner wall of the chamber was a leading cause of the formation of cubic and hexagonal stacking sequences during deposition at 2.2 × 10$^{-6}$ Pa (Figs. 3 and 5). The formation of amorphous water at about 2.2 × 10$^{-5}$ Pa also supports the validity of Eqs. 3 and 4, because an increase of the gas flux led to a smaller $t_{cover}$ than that at 2.2 × 10$^{-6}$ Pa.

From Eq. 4, the upper limit of $D_s$ of water molecules for the formation of amorphous water can be derived as follows:

$$D_s < \frac{A_{ice}}{t_{cover}}, \tag{5}$$

Adopting $A_{ice} = 8.8 \times 10^{-16}$ cm$^2$ and $t_{cover} = 147 \pm 10$ s gives an upper limit of $D_s \leq 6.0 \pm 0.4 \times 10^{-18}$ cm$^2$ s$^{-1}$. This value is within error of the value of $D_s$ for the surface of amorphous water at 120 K ($1.1^{+4.9}_{-0.9} \times 10^{-17}$ cm$^2$ s$^{-1}$), as calculated using the activation energy (13.7 ± 1.7 kJ mol$^{-1}$) and preexponential factor (1.1 × 10$^{-11}$ cm$^2$ s$^{-1}$) reported by Jung et al.[52] Although we cannot precisely determine $D_s$ in this study, the change in ice structure from amorphous to ice $I_c$ at $l$ = 13–17 nm suggests that $D_s$ lies close to the boundary between the conditions of formation of amorphous and crystalline ice (Eq. 6), and it may increase with increasing ice thickness:

$$D_s \approx \frac{A_{ice}}{t_{cover}} \tag{6}$$

A plausible reason for the increase in $D_s$ on amorphous water is the change in surface morphology with increasing ice thickness. Thürmer and Bartelt used scanning tunneling microscopy (STM) to study the thickness dependence of the morphology of amorphous water on Pt(111) at 100 K,[53] finding that the effect of the Pt substrate on the ice morphology diminishes with increasing thickness of the ice up to 6 nm. The estimated lateral scale of surface irregularities is 5–10 nm in amorphous water of 6 nm thickness.[53,96] Tomaru et al. used atomic force microscopy (AFM) to find a lateral scale of surface irregularities of 10–30 nm for 13.5 nm-thick amorphous water on a Si(111) 7 × 7 substrate at 100 K, indicating that this amorphous water had a laterally smooth surface compared with the 6 nm-thick amorphous water.[96] Non-contact AFM observations by Donev et al. of 14 nm-thick amorphous water on Au(111) at 108 K revealed small circular inhomogeneities typically 1–4 nm in height and 20–30 nm wide on the surface of the amorphous water,[97] similar to the results of Tomaru et al.[96] Based on these previous STM and AFM observations,[53,96,97] we speculate that the $D_s$ on amorphous water with $l$ = 13–17 nm can become sufficiently high to enable the efficient nucleation of ice $I_c$, because water molecules should be able to diffuse efficiently on a laterally smooth surface. A large smooth area should be also advantageous to the growth of ice $I_c$ after nucleation.

Previous investigations of the formation conditions of ice $I_c$ and/or ice $I_{sd}$ have considered mainly homogeneous nucleation systems.[33,98–106] For example, Johari theoretically proposed that water droplets smaller than 30 nm and flat films of water thinner than 10 nm freeze to ice $I_c$ at 160–220 K rather than ice $I_h$ because of the difference in the solid–liquid interfacial free energy ($\gamma$) of the respective ice–liquid interfaces.[99] Zhang et al. later calculated a critical size of about 10 nm for water droplets to freeze into ice $I_c$ by considering the effects of the temperature dependence of $\gamma$ and solid–liquid interface stresses.[100]

The Debye–Scherrer rings observed in the present study (Fig. 3) indicate the existence of randomly oriented small ice $I_c$ crystallites of several nanometers (typically about 2.5 nm)[107] on the amorphous water surface at 120 K. Compared with homogeneous nucleation, although much less is known about heterogeneous nucleation of water, especially at low temperatures,[108,109] the nucleation and growth of nanometer-sized ice $I_c$ is qualitatively consistent with the thermodynamic calculations of Johari and Zhang et al.[99,100]

Figure 5 shows that a cubic stacking sequence is dominant during deposition for up to 240 min ($l$ = 36 nm), whereas hexagonal stacking appears at vapor deposition durations of 300–420 min ($l$ = 45–63 nm). This shows that the nucleation and growth of ice $I_h$ requires the growth of ice $I_c$ crystallites. Although clarifying the microscopic mechanism of the formation of ice $I_h$ on ice $I_c$ is beyond the scope of this study, we discuss the following mechanisms: epitaxial-mediated 2D nucleation and double spiral growth.

As the (111) planes of ice $I_c$ are identical to the (0001) planes of ice $I_h$,[33] heterogeneous 2D nucleation of the hexagonal phase can occur on the planes of a cubic crystal during vapor deposition in addition to the growth of ice $I_c$.[110] Thürmer and Nie reported another mechanism whereby double spirals provide a kinetic route of growing ice $I_h$ on ice $I_c$ without nucleating new layers.[111] Their STM and AFM observations of vapor-deposited ice on Pt(111) at 140 K revealed that the crystal structure switched from ice $I_c$ to ice $I_h$ at a thicknesses of about 20 nm. They observed S-shaped double spirals on the surfaces of crystalline ice of 20–40 nm thickness, whereas only single spirals were observed for ice thinner than 15 nm. The authors proposed that the single spiral growth produced ice of uniform intralayer stacking (i.e., ice $I_c$) and that the S-shaped double spiral growth on cubic ice generated ice with alternating intralayer stacking, which resulted in ice $I_h$.[111]

The epitaxial-mediated 2D nucleation of ice $I_h$ competes with the growth of the preexisting ice $I_c$. If the ice $I_c$(111) surface is rough, water molecules deposited on the surface are immediately incorporated into the ice $I_c$ crystal via kinks, and the 2D nucleation of ice $I_h$ is hardly expected.[110] A smooth ice $I_c$(111) surface is thus desirable for epitaxial-mediated 2D nucleation. For double spiral growth, Thürmer and Nie proposed that screw dislocations with a double Burgers vector (i.e., a vector twice as long as the interlayer spacing) are needed to create the S-shaped double spirals.[111] Clarification of the contribution of each of the above mechanisms to the present experimental results requires microscopic observations of the ice surface morphology, with particular focus on the surface roughness, including kinks and screw dislocations.

The present *in situ* RHEED and IRRAS study clearly demonstrates that the structure of vapor-deposited ice is sensitive not only to the substrate temperature and vapor pressure, but also to the ice thickness. As a result, the vapor-deposited ice prepared by 420 min with $l = 63 \pm 4$ nm is a mixture of amorphous water and ice $I_c$ and ice $I_h$ (i.e., ice $I_{sd}$; Figs. 4A and 5H). Thus, the present results suggest that amorphous water and cubic and hexagonal stacking sequences can coexist in low-temperature (around 120 K) regions of the mesosphere.[3,8,112] A more accurate discussion of the structures of ices in polar mesospheric clouds would benefit from a systematic experimental study over a wide range of temperatures (100–145 K) and vapor pressures ($10^{-7}$–$10^{-5}$ Pa). Further research is ongoing to understand the conditions required for the formation of amorphous water and ices $I_h$, $I_c$, and $I_{sd}$ by vapor deposition.

The present study also provides fundamental insights into the nucleation and growth of solid materials.[113–116] The three-step growth of ice (i.e., amorphous water → ice $I_c$ → ice $I_h$) during vapor deposition indicates that the formation mechanism of vapor-deposited ice cannot be explained by the simple classical theory for heterogeneous nucleation assuming a single-step formation of stable ice $I_h$ in the bulk phase.[117] Given the negligible bulk structural transformation at 120 K, the surfaces of amorphous water and ice $I_c$ serve as precursors for the formation of ice $I_c$, and ice $I_h$, respectively.[116] The empirical Ostwald rule of stages used in crystallography states the following: "When leaving a given state and in transforming to another state, the state which is sought out is not the thermodynamically stable one, but the state nearest in stability to the original state."[118] This rule originated from observations of a wide range of solution-based crystallization systems,[110,119–121] but there has been no proof of its general validity, and the broadness of its applicability is not well explored.[120] This study estimated the saturation ratio during vapor deposition to be $8.8 \times 10^3$, considering the ratio of the experimental vapor pressure ($2.2 \times 10^{-6}$ Pa) to the calculated equilibrium vapor pressure of ice $I_h$ at 120 K ($2.5 \times 10^{-10}$ Pa),[122] which is much higher than that in typical solution systems. The sequential formation of amorphous water, ice $I_c$, and ice $I_h$ reported here shows that the Ostwald rule of stages appears applicable to water under such conditions, far from equilibrium.

**Experimental Methods**

The experimental apparatus consisted of an ultrahigh-vacuum chamber, an aluminum (Al) substrate mounted on the cold head of a closed-cycle helium (He) refrigerator, a gas doser, an electron gun, a fluorescent screen, and a Fourier transform infrared spectrometer (FT-IR) (Fig. 1).[46,47] The vacuum chamber was evacuated to the base pressure of $10^{-8}$ Pa using a turbo molecular pump (nEXT300D, Edwards). A mirror-polished Al alloy 2017 substrate (38 mm diameter), including its amorphous native oxide, was set on a copper (Cu) substrate holder (Fig. S6 in the Supporting Information). The Cu sample holder was connected to the cold head of a closed-cycle He refrigerator (RDK-101D, Sumitomo Heavy Industries) and installed at the center of the vacuum chamber using a laboratory-made bore-through rotary feedthrough. The temperature of the Al substrate was measured using a silicon (Si) diode sensor (DT-670, Lakeshore) placed on the back side of the copper sample holder and controlled using a temperature controller (Model 325, Lakeshore) and a 40 W ceramic heater (MC1010, Sakaguchi E. H Voc Corp.).

Vapor-deposited ice samples were prepared using $H_2O$ with 3.5 mol% HDO. Purified $H_2O$ (resistivity ≥ 18.2 MΩ cm at 298 K) from a Millipore Milli-Q water purification system was mixed with 2.0 wt% (1.8 mol%) $D_2O$ (deuteration degree > 99.9%; Merck, USA) to obtain a water sample containing about 3.5 mol% HDO with a negligible amount of $D_2O$. Partial deuteration facilitated measurement of the OD stretching vibration of HDO molecules in the bulk ice decoupled from intramolecular and intermolecular OH stretching vibrations.[72,73] The decoupled OD stretching band at around 2400 cm$^{-1}$ was spectrally isolated and off-resonance with the major OH stretching band between 3800 and 3000 cm$^{-1}$, and the concentration of HDO (3.5 mol%) was sufficiently low for intermolecular vibrational coupling to be ignored. The water sample was degassed by several freeze–pump–thaw cycles before deposition. Water ice was produced on the Al substrate at typically 120 K by background water deposition through a variable leak valve (Fig. 1). The pressure during vapor deposition was measured as $5.0 \times 10^{-7}$ Pa using a cold cathode gauge (IKR270, Pfeiffer). As described in the main text, the actual pressure around the Al substrate was estimated to be $2.2 \pm 0.2 \times 10^{-6}$ Pa, which was 4–5 times that measured by the cold cathode gauge ($5.0 \times 10^{-7}$ Pa). As the variable leak valve for vapor deposition was located in front of the Al substrate (Fig. 1), the vapor pressure around the Al substate was likely higher than that around the cold-cathode gauge.

IRRAS measurements were performed using an FT-IR spectrophotometer (FT/IR-6600, JASCO Corporation, Japan). Fourier-transform-modulated and unpolarized IR light was directed from the FT-IR spectrophotometer to the Al substrate in the vacuum chamber with an incident angle ($\varphi$) of 37.5°. The reflected IR light was detected with a type II superlattice IR detector (P15409-901, Hamamatsu, Japan). The accumulation number of the interferogram collection (single-beam measurement) was 1000 (430 s), with a spectral resolution of 4 cm$^{-1}$ at a scan speed of 8 mm s$^{-1}$. Errors in ice thickness were assessed from three independent measurements.

The structure of vapor-deposited ice was examined in situ by RHEED. An electron beam (20 keV) generated by an electron gun (RDA-004G, R-DEC Co., Ltd.) was impinged on the ice surface with an incident angle ($\theta$) of 2°–3°.[46] The kinetic energy of these electrons corresponded to a de Broglie wavelength, $\lambda$, of about 0.09 Å. Considering the inelastic mean free path of electrons ($L$) in liquid water (about 50 nm at 20 keV),[123] the penetration depth of electrons was estimated to be approximately 2–3 nm using $L \sin\theta$. Hence, low-temperature RHEED probed the top several layers of the ice samples. The resulting RHEED patterns were projected onto a fluorescent screen and photographed using a digital camera (α7RII, Sony) with a macro lens (SEL50M28, Sony). RHEED measurements were performed as quickly as possible (typically within 1 min) to prevent the electron beam damaging the ice samples. We confirmed no discernible change in the RHEED patterns, even after prolonged electron irradiation for 10 min.

ASSOCIATED CONTENT

**Supporting Information**.

The following files are available free of charge.

Literature values for complex refractive index ($n + ik$) and calculated longitudinal optic energy-loss function ($f_{LO}$) of amorphous water at 70 K (Table S1 and Figure S1); literature values of complex refractive index ($n + ik$) and calculated longitudinal optic energy-loss function ($f_{LO}$) of polycrystalline ice I at 150 K (Table S2 and Figure S1); RHEED pattern of reference ice I at 120 K (Figure S2); summary of Gaussian fitting for the decoupled-OD stretching vibration band in vapor-deposited ices at 120 K (Table S3); RHEED patterns of vapor-deposited ices after annealing at 120 K (Figures S3 and S4); RHEED patterns of vapor-deposited ice at 120 K after deposition for 120 min at about $2.2 \times 0.2 \times 10^{-5}$ Pa (measured pressure of $5.0 \times 10^{-6}$ Pa) (Figure S5); RHEED pattern of Al substrate at room temperature (Figure S6).

AUTHOR INFORMATION

**Corresponding Author**

*Tetsuya Hama

Email: hamatetsuya@g.ecc.u-tokyo.ac.jp

ORCID numbers

Reo Sato: 0009-0009-4600-3243

Hiroyuki Koshida: 0000-0002-2416-569X

Atsuki Ishibashi: 0000-0001-9751-6641

Tetsuya Hama: 0000-0002-4991-4044**Notes**

The authors declare no competing financial interest.

ACKNOWLEDGMENT

This work was supported by grants from the Japan Society for the Promotion of Science KAKENHI (Nos. 24H00264, and 23H03987), the Kurita Water and Environment Foundation (No. 23D004), the Sumitomo Foundation (fiscal 2023 grant for Basic Science Research Projects, No. 2300811), and the JST FOREST Program (No. JPMJFR231J).